\documentclass[pra,aps,floats,letterpaper,floatfix,footinbib,superscriptaddress,preprintnumbers]{revtex4}
\usepackage{amsmath}
\usepackage{color}
\usepackage[mathenv]{}
\usepackage{graphicx}
\usepackage{ulem} 
\usepackage{natbib}
\begin{document}

\title{A Quantum-Enhanced Prototype Gravitational-Wave Detector}
\author{K. Goda}
\affiliation{LIGO Laboratory, Massachusetts Institute of Technology, Cambridge, Massachusetts 02139 USA}
\author{O. Miyakawa}
\affiliation{LIGO Laboratory, California Institute of Technology, Pasadena, California 91125 USA}
\author{E. E. Mikhailov}
\affiliation{The College of William $\&$ Mary, Williamsburg, Virginia 23187 USA}
\author{S. Saraf}
\affiliation{Rochester Institute of Technology, Rochester, New York 14623 USA}
\author{R. Adhikari}
\affiliation{LIGO Laboratory, California Institute of Technology, Pasadena, California 91125 USA}
\author{K. McKenzie}
\affiliation{Center for Gravitational Physics, The Australian National University, ACT 0200, Australia}
\author{R. Ward}
\author{S. Vass}
\author{A. J. Weinstein}
\affiliation{LIGO Laboratory, California Institute of Technology, Pasadena, California 91125 USA}
\author{N. Mavalvala}
\affiliation{LIGO Laboratory, Massachusetts Institute of Technology, Cambridge, Massachusetts 02139 USA}

\begin{abstract}
The quantum nature of the electromagnetic field imposes a
fundamental limit on the sensitivity of optical precision
measurements such as spectroscopy, microscopy, and interferometry
\cite{scully1997cambridge}. The so-called quantum limit is set by
the zero-point fluctuations of the electromagnetic field, which
constrain the precision with which optical signals can be measured
\cite{breitenbach1997nature,walls1983nature,caves1981prd}. In the
world of precision measurement, laser-interferometric gravitational
wave (GW) detectors \cite{abramovici1992science,caves1981prd,adligo}
are the most sensitive position meters ever operated, capable of
measuring distance changes on the order of $10^{-18}$ m RMS over
kilometer separations caused by GWs from astronomical sources
\cite{thorne1987cambridge}. The sensitivity of currently operational
and future GW detectors is limited by quantum optical noise
\cite{adligo}. Here we demonstrate a 44\% improvement in
displacement sensitivity of a prototype GW detector with suspended
quasi-free mirrors at frequencies where the sensitivity is shot-noise-limited, by injection of a squeezed state of light \cite{scully1997cambridge,breitenbach1997nature,walls1983nature}. This
demonstration is a critical step toward implementation of
squeezing-enhancement in large-scale GW detectors.
\end{abstract}

\maketitle

\section{Introduction}
Laser-interferometric GW detectors, such as those of the Laser
Interferometer Gravitational-Wave Observatory (LIGO)
\cite{abramovici1992science}, are designed to measure distance
changes on the order of $10^{-18}$ m, or one-thousandth the diameter
of the proton, caused by GWs from astronomical sources such as
neutron star or black hole binaries, supernovae, and the Big Bang
\cite{thorne1987cambridge}. Their goals are to further verify
Einstein's theory of general relativity and open an entirely new
window onto the universe
\cite{thorne1987cambridge,abramovici1992science}. The sensitivity of
currently operational GW detectors is largely limited by quantum
optical noise. Next-generation GW detectors, such as Advanced LIGO
\cite{adligo}, planned to be operational in the next few years, are
anticipated to be limited in sensitivity by quantum optical noise at
almost all frequencies in the GW detection band (10 Hz -- 10 kHz).
The mirrors used in these interferometric GW detectors are suspended
as pendulums, serving as quasi-free test masses in the GW detection
band \cite{abramovici1992science}. The quantum noise arises from
uncertainty products associated with the commutation relations
between conjugate field operators.

The quantum limit can be circumvented by use of nonclassical or
\textit{squeezed} states of light
\cite{scully1997cambridge,breitenbach1997nature,walls1983nature},
where fluctuations are reduced below the symmetric quantum limit in
one quadrature at the expense of increased fluctuations in the
canonically conjugate quadrature. Squeezed states comprise a
phase-dependent distribution of zero-point fluctuations such that
the fluctuations in one quadrature are smaller than those of a
coherent state, at the expense of increased fluctuations in the
canonically conjugate quadrature, while preserving the Heisenberg
limit on the uncertainty product. Since squeezed states were first
observed by Slusher \textit{et al.} in 1985 \cite{slusher1985prl}, a
number of experimental efforts have realized the proof-of-principle
of quantum noise reduction or squeezing-enhancement in various
high-precision applications, such as spectroscopic measurement
\cite{polzik1992prl}, lateral displacement measurement and imaging
\cite{treps2002prl,treps2003science}, polarization measurement
\cite{grangier1987prl}, and interferometric phase measurement
\cite{xiao1987prl}. Incorporating the technique of
squeezing-enhancement into practical devices has remained a great
challenge, either because the quantum limit is not reached due to
excess classical noise, or because it is less onerous to enhance
sensitivity by optimizing other parameters classically. Squeezed
states are a useful ingredient for quantum teleportation
\cite{furusawa1998science}, quantum cryptography
\cite{grosshans2002prl}, and quantum lithography \cite{boto2000prl},
but these applications are yet to reach the stage of practical
implementation, in part due to the complexity and technical
challenges of working with squeezed states.

Laser-interferometric GW detectors such as LIGO
\cite{abramovici1992science}, VIRGO \cite{VIRGO}, GEO600
\cite{GEO600}, and TAMA300 \cite{TAMA300} are so sensitive (they
measure distance changes of $10^{-18}$ m over kilometer separations)
that they have already confronted the quantum limit
\cite{ligoS5instr}. In next-generation detectors, such as Advanced
LIGO \cite{adligo}, optimization of classical parameters will reach
the limits of technology. Any further improvement in sensitivity
must rely on quantum techniques such as squeezing-enhancement,
making GW detectors an important practical application of squeezed
states of light.

The quantum nature of light reveals itself in two effects that limit
the precision of an optical measurement of mirror position in
laser-interferometric GW detectors: (i) \textit{photon shot noise},
which typically dominates at frequencies above 100 Hz, arises from
quantum uncertainty in the number of photons at the interferometer
output; and (ii) \textit{quantum radiation pressure noise},
typically dominant at frequencies below 100 Hz, that arises from
mirror displacements induced by quantum radiation pressure
fluctuations \cite{kimble2002prd,adligo}. Both effects are caused by
quantum fluctuations of a vacuum electromagnetic field that enters
the antisymmetric port of the interferometer \cite{caves1981prd}.
The displacement noise associated with the shot noise and quantum
radiation pressure noise of a simple Michelson interferometer on a
dark fringe in the frequency domain is given by
\cite{abramovici1992science}
\begin{eqnarray}
\Delta x_{\rm shot} = \sqrt{\frac{\hbar c \lambda}{\pi P}} \hspace{0.5cm}
{\rm and} \hspace{0.5cm} \Delta x_{\rm rad} = \sqrt{\frac{\hbar P}{\pi^3
c\lambda m^2 f^4}}, \label{eq:quantnoise}
\end{eqnarray}
where $c$ is the speed of light in vacuum, $\hbar$ is Planck's
constant, $\lambda$ is the laser wavelength, $m$ is the (reduced)
mass of the mirrors, $f$ is the measurement frequency, and $P$ is
the optical power incident on the beamsplitter.

The quantum limit in the laser-interferometric GW detector can be
overcome by the injection of squeezed states of light into the
antisymmetric port of the interferometer \cite{caves1981prd}.
Following the 1981 proposal of Caves \cite{caves1981prd} to improve
the sensitivity of quantum-noise-limited laser interferometers by
squeezed state injection, a handful of experimental efforts have
realized the proof-of-principle on the table-top scale at MHz
frequencies \cite{xiao1987prl,mckenzie2002prl,vahlbruch2005prl}. Our
demonstration of squeezing-enhancement shows improved sensitivity in
a suspended-mirror prototype GW detector by injecting a squeezed
vacuum field with an {\it inferred} level of 9.3$\pm$0.1 dB relative
to shot noise into the antisymmetric port (see Methods). An
important distinction between our experiment and previous efforts is
that it is the first implementation of squeezing-enhancement in a
prototype GW detector with suspended optics and a control and
readout scheme similar to those used in the currently operational
LIGO detectors, making it necessary to confront dynamical effects
introduced by suspended mirrors such as optical springs
\cite{miyakawa2006prd}. It is, therefore, a critical step toward
implementation of squeezing-enhancement in large-scale GW detectors.
In all these experiments, including the one reported here, quantum
radiation pressure noise was buried under other technical noise
sources such as seismic noise and mirror thermal noise, and only the
shot noise limit was accessible (not buried under technical noise)
for showing squeezing-enhancement. As of today, quantum radiation
pressure noise has not been observed in any experimental setting.

Terrestrial GW detectors typically comprise a Michelson
interferometer with a Fabry-Perot cavity in each arm, to increase
the phase sensitivity of the detector. The Michelson interferometer
is operated on or near a dark fringe. Since most of the incident
light returns toward the laser source, the GW-induced signal can be
increased by recycling the laser power back toward the beamsplitter.
This is achieved by placing a partially transmitting mirror -- the
``power recycling'' mirror -- between the laser source and the
beamsplitter. Typical power recycling gains of 30 to 70 have be
realized in presently operational GW detectors. Similarly, a
partially transmissive mirror can also be placed at the
antisymmetric port of the beamsplitter to further enhance the
GW-induced signal at the interferometer output. This ``signal
recycling'' mirror forms a complex optical cavity with the rest of
the interferometer. The frequency-dependent optical response of the
detector to incident GWs can be tuned by operating the signal
recycling cavity at various detunings from resonance. Signal
recycling is utilized in the GEO600 detector \cite{GEO600}, and is
planned for the Advanced LIGO detector \cite{adligo}.

\begin{figure}[t]
\includegraphics[angle=0, width=0.7\columnwidth]{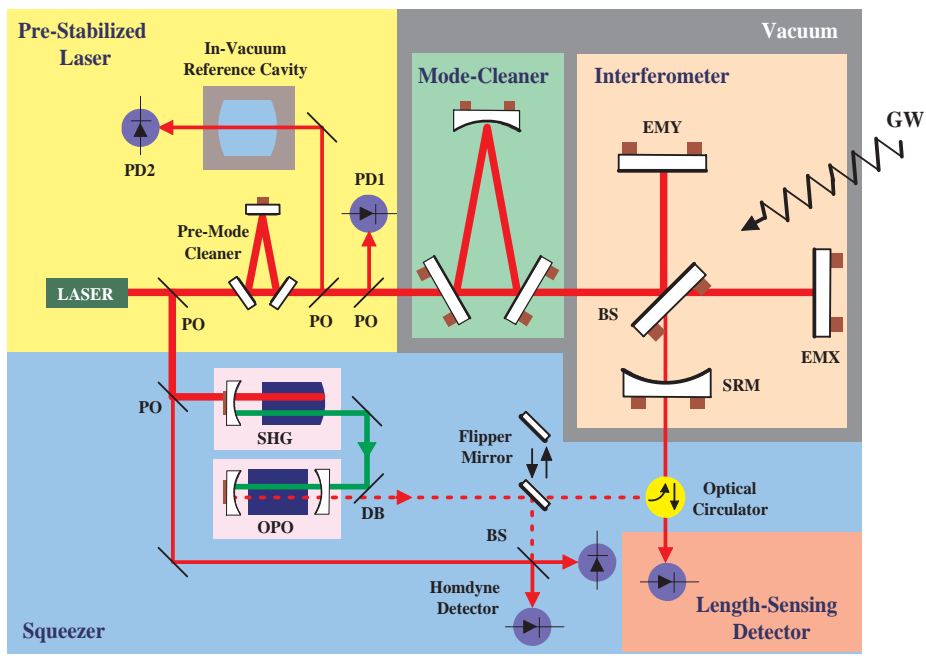}
\caption{Schematic of the quantum-enhanced prototype gravitational-wave (GW) detector that
consists of five parts. BS: 50/50 beamsplitter; PD1 and PD2:
photodetectors; PO: pickoff beamsplitter; DB: dichroic beamsplitter; SRM: signal-recycling mirror; EMX/EMY:
end mirrors along the x/y axis, respectively; SHG: second-harmonic
generator; OPO: optical parametric oscillator; GW: gravitational wave. The SHG, OPO,
reference cavity, pre-mode cleaner, mode-cleaner, and
signal-recycling cavity are are all locked by adaptations of Pound-Drever-Hall locking \cite{drever1983applphysb}. PD1 and PD2 are used for the laser intensity and frequency
stabilization while two photodetectors in the interferometer (not shown) and the length sensing detector are used to control the
interferometer. All the mode-cleaner and interferometer optics are
suspended by single loop pendulums.}
\label{apparatus}
\end{figure}

\section{Experiment}
The experiment reported here used a sub-configuration of the
complete Advanced LIGO interferometer -- a signal-recycled Michelson
interferometer (SRMI), chosen in part because it is an important new
feature of the optical configuration envisioned for Advanced LIGO
\cite{adligo,miyakawa2006prd}. The shot noise limited displacement
sensitivity of the SRMI is given in the frequency domain by
\cite{goda2007mit}
\begin{eqnarray}
\Delta x_{\rm SRMI} = \frac{1}{\sqrt{|G|}}\sqrt{\frac{\hbar c \lambda}{\pi
\eta P}}e^{-R},
\end{eqnarray}
where $G$ is the signal-recycling gain, given by $G=[t_s/(1-r_s r_m
e^{-2i\phi})]^2$, $\eta$ is the power transmission efficiency from
the signal-recycling mirror to the photodetector (including the
quantum efficiency of the photodetector), and $R$ is the squeeze
factor \cite{walls1983nature}. Here $r_s$ and $t_s$ are the amplitude reflectivity and
transmissivity of the signal-recycling mirror, $r_m$ is the
reflectivity of the Michelson interferometer, and $\phi$ is the
signal-recycling cavity detuning. We note that setting $G=1$ and
$R=0$ leads to the familiar expression for the shot noise limit in a
Michelson interferometer, given in Eq. \ref{eq:quantnoise}. In the
experiment described below, $\eta, r_s$, and $r_m$ are measured to
be 0.825, $\sqrt{0.925}$, and $\sqrt{0.995}$ respectively, and $P$
and $\phi$ are obtained from fits to be 57 mW and 0, respectively.

\begin{figure}[t]
\includegraphics[width=0.7\textwidth]{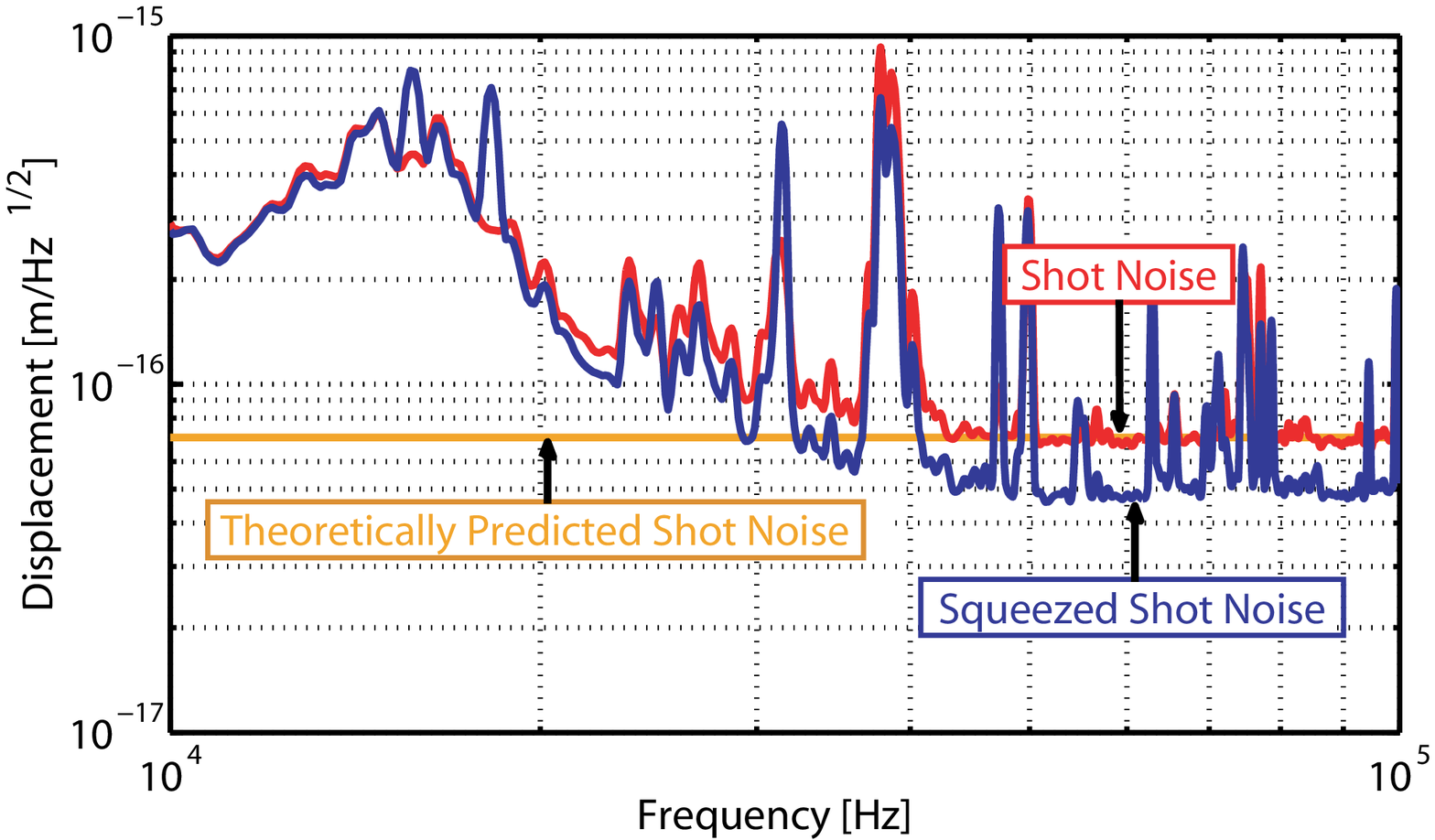}
\caption{The noise floor of the signal-recycled
Michelson interferometer with (blue) and without (red) the injection
of squeezed vacuum. The theoretically predicted shot noise level
based on the measured optical power is also shown (orange). The
interferometer is shot-noise-limited at frequencies above 42 kHz.
Injection of squeezing leads to broadband reduction of the shot
noise in the shot-noise-limited frequency band.}
\label{SRMI}
\end{figure}

This experiment was carried out in a prototype GW detector with
suspended mirrors \cite{miyakawa2006prd}, designed to closely mimic
the kilometer-scale LIGO detectors. A schematic of the experiment is
shown in Fig. 1. Its major components are (i) an intensity- and
frequency-stabilized Nd:YAG Master Oscillator Power Amplifier (MOPA)
laser with a throughput of 5 W at 1064 nm that serves as the light
source for both the interferometer and the squeezed vacuum
generator; (ii) a triangular optical cavity -- or mode cleaner --
which consists of three free hanging mirrors with a linewidth of 4
kHz to further stabilize the intensity, frequency, and mode of the
laser; (iii) a test interferometer configured as a SRMI, comprising
a 50/50 beamsplitter, two high-reflectivity end mirrors, and a
signal-recycling mirror, all suspended as single loop pendulums
(causing the optics to behave as inertial free masses); (iv) a
squeezed vacuum generator -- or squeezer -- that consists of a
second-harmonic generator (SHG), a sub-threshold vacuum-seeded
optical parametric oscillator (OPO) pumped by a second-harmonic
field, a monitor homodyne detector, and an optical circulator to
inject the generated squeezed vacuum field to the interferometer;
and (v) a high quantum efficiency photodetector to sense
differential motion of the interferometer mirrors. The Michelson
interferometer is locked on a dark fringe using a static
differential offset such that a small amount of the carrier light
exits the signal-recycling cavity, while the signal-recycling cavity
is locked on a carrier resonance. This DC component of the carrier light at the antisymmetric port acts as a local oscillator field for a GW-induced signal to beat against, forming a homodyne detection or ``DC readout'' scheme \cite{adligo}.

The noise performance of the interferometer is shown in Fig. 2. The
comparison between the measured noise floor without squeezing and
the theoretically predicted noise floor based on the measured
optical power of 100 $\mu$W indicates that the interferometer is
shot-noise-limited at frequencies above 42 kHz. In addition, the
interferometer output power is changed by adjusting the Michelson
offset to verify the $\sqrt{P}$ scaling of the shot noise spectral
density. At frequencies below 42 kHz, the noise is dominated by
laser intensity noise and uncontrolled length fluctuations of the
interferometer. The peaks at frequencies above 42 kHz are also due
to the interferometer length fluctuations. In the shot-noise-limited
frequency band, the detector sensitivity is $(6.9 \pm 0.1)\times10^{-17}$
m/$\sqrt{\rm Hz}$. Systematic uncertainty in the displacement
calibration is estimated to be 10\%, but does not affect the relative improvement
achieved by squeeze injection that was observed.

\section{Results}
The result of the squeezing-enhancement in the interferometer is
also shown in Fig. 2. The comparison between the two spectra shows
that the noise floor of the interferometer was reduced by the
injection of the squeezed vacuum field in the shot-noise-limited
frequency band. Fig. 3 shows the noise floor with a simulated GW
signal at 50 kHz, with and without injected squeezing. The broadband
quantum noise floor was reduced from $(6.9 \pm 0.1)\times10^{-17}$
m/$\sqrt{\rm Hz}$ to $(4.8 \pm 0.1)\times10^{-17}$ m/$\sqrt{\rm
Hz}$, while the strength of the simulated GW signal was retained.
This corresponds to a 44\% increase in signal-to-noise ratio (SNR)
or detector sensitivity. In kilometer-scale GW detectors, this would
correspond to a factor of $1.44^3 = 3.0$ increase in detection rate
for isotropically distributed GW sources. The squeeze factor is
found to be $R=0.36\pm 0.03$ where the error comes from the
variance in each noise floor. Other peaks in the squeezing spectrum
are due to optical crosstalk between the interferometer and OPO,
arising from inadequate isolation of the OPO from the interferometer
output. It can be completely resolved by improved Faraday isolation,
or use of an OPO in a bow-tie cavity configuration that
geometrically separates the two fields \cite{grosse2006prl}, and is
not expected to be a problem in operational GW detectors.

The measurable squeezing effect was limited to frequencies above 42
kHz in this experiment, since the quantum noise is masked by
classical noise at lower frequencies. Many of the factors that led
to this limit in the experiment presented here are not expected to
effect squeezing-enhancement in the LIGO interferometers. One of the
greatest challenges of using suspended mirrors -- a crucial feature
of this work -- is the coupling of seismic noise to the detector
output, due to large fluctuations in the positions and angles of the
mirrors. In addition to the seismic noise coupling, laser frequency
and intensity fluctuations also degrade the low frequency
performance of the detector. Unlike the urban campus siting of the
prototype detector used here, the LIGO detectors are located in
remote sites with much quieter seismic and acoustic environments,
and are equipped with better vibration isolation and mirror control
systems. Moreover, in the kilometer-scale detectors, the long Fabry-Perot cavities act as optical filters to mitigate laser noise in the
GW band \cite{ligoS5instr}. Other challenges associated with
suspended-mirror interferometry that had to be overcome in the
present experiment include control of the mode overlap between the
(relatively static) squeezed input and (dynamically fluctuating)
interferometer output fields, interfacing of the in-vacuum parts of
the experiment (the interferometer) with external optical systems
(the squeezer), and optical isolation to mitigate optical feedback.
Consequently, we expect the squeezing-enhancement to be effective at
lower frequencies within the GW detection band when implemented on
the long baseline LIGO detectors. This implementation of
squeezing-enhancement in a GW detector prototype with the suspended
optics and the readout and control scheme similar to those used in
the currently operational LIGO detectors firmly establishes the
practical feasibility of squeezing injection for future improvements
to existing GW detectors worldwide.

\begin{figure}[t]
\includegraphics[width=0.7\textwidth]{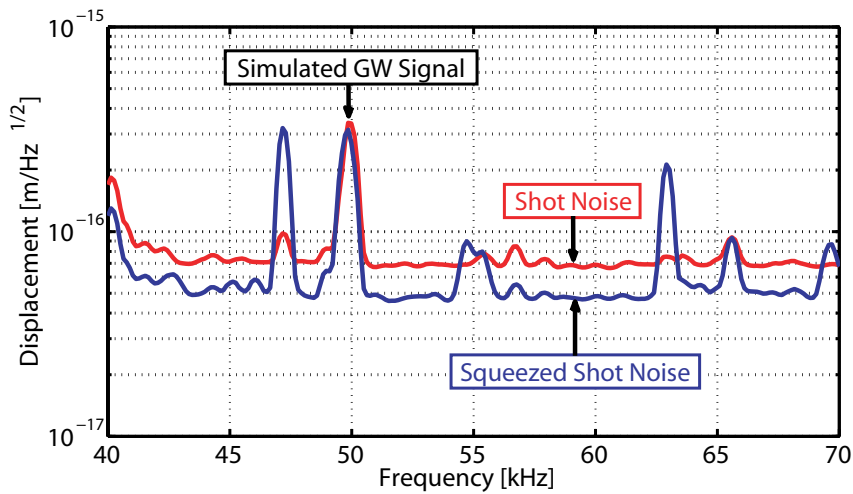}
\caption{A zoomed graph of the noise floor of the
interferometer with a simulated gravitational-wave (GW) signal at 50 kHz with (blue) and
without (red) the injection of squeezed vacuum. The simulated GW was generated by exciting the beamsplitter of the interferometer. The broadband shot
noise floor was reduced from $(6.9 \pm 0.1)\times10^{-17}$ m/$\sqrt{\rm Hz}$ to $(4.8 \pm 0.1)\times10^{-17}$
m/$\sqrt{\rm Hz}$ by the injection of squeezing while the strength of the simulated
GW signal was retained, resulting in a 44\% increase in signal-to-noise
ratio or detector sensitivity.}
\label{SRMIm}
\end{figure}

\section{Experimental Details}
Each optic of the interferometer is suspended as a single loop
pendulum mounted on a passive vibration isolation system within a
single vacuum volume with a pressure of $10^{-6}$ torr. At
frequencies above the pendulum resonant frequency ($\sim1$ Hz), the
suspensions attenuate seismic noise, causing the optics behave as
inertial free masses. The mirrors are held in place by feedback
control designed to suppress low-frequency seismically driven
motion. Magnets affixed to each optic are surrounded by
current-carrying coils that actuate on the mirror positions and
angles. Adaptations of the Pound-Drever-Hall locking \cite{drever1983applphysb} are used
to keep the mirrors of the interferometer at the desired operating
point. More details can be found in Refs. \cite{abbott2001LIGOtechnote,miyakawa2006prd}.

The OPO, operated below threshold, is used to produce a squeezed
state by correlating the upper and lower quantum sidebands centered
around the carrier frequency, in the presence of an energetic pump
field. Since all GW detectors presently use high power Nd:YAG lasers
sources at 1064 nm, generating squeezed states at 1064 nm is
essential. The OPO is a cavity composed of a 10 mm long quasi-phase
matched periodically-poled KTiOPO$_4$ crystal with anti-reflection
coated flat surfaces and two external coupling mirrors
\cite{goda2008optlett}. To generate a low-frequency squeezed state,
the OPO is vacuum-seeded and pumped by 320 mW of a second-harmonic
field which is generated by the SHG from the same laser source that
is incident on the interferometer. An auxiliary laser that is
frequency-shifted by 642 MHz relative to the carrier frequency is
used to lock the OPO cavity by using a PZT on one of the coupling
mirrors. This frequency-shifted light is orthogonally polarized to
the vacuum field that seeds the OPO cavity and to the pump field,
and therefore, does not enter the interferometer. This type of OPO
locking scheme is necessary to generate a squeezed state of
\textit{vacuum} which is squeezed at all frequencies, not a squeezed
state of \textit{light} which is typically squeezed only at MHz
frequencies due to laser excess noise at frequencies below 1 MHz
\cite{vahlbruch2005prl, goda2007mit}.

The balanced homodyne detector is used to monitor the generated
squeezed vacuum field before injection to the interferometer. It is
composed of a 50/50 beamsplitter and a pair of photodiodes with
matched quantum efficiencies of 93$\%$. With a mode-cleaned
coherent local oscillator field, the homodyne efficiency of this
readout is 99.2$\%$. The squeezing level measured by this detector
is $7.4 \pm 0.1$ dB \cite{goda2008optlett}. Based on this value and
the composite detection efficiency, the squeezing level at the
output of the OPO is inferred to be 9.3$\pm$0.1 dB
\cite{goda2007mit,goda2008optlett}.

The squeezed vacuum field is injected into the antisymmetric port of
the interferometer via an optical circulator, a mode-matching
telescope, and steering mirrors. To ensure a high coupling
efficiency of the squeezed vacuum field to the interferometer field,
the transmission of the interferometer output through the OPO cavity
in a TEM$_{00}$ mode with the interferometer locked on a bright
fringe is optimized by using the steering mirrors and mode-matching
telescope. With the OPO cavity length scanned, the mode structure of
the OPO transmission indicates the coupling efficiency of the
interferometer mode to the OPO cavity mode or equivalently the
optical loss due to the mode-mismatch between the interferometer and
squeezed vacuum fields.

The interferometer output is detected by the length sensing
photodetector with a quantum efficiency of 93$\%$. To show the
validity of squeezing, it is critical to calibrate the shot noise
level correctly. Since the Michelson interferometer and
signal-recycling cavity are locked by using RF sidebands (as well as
the DC carrier light at the length sensing detector), it is
important to verify that the carrier-to-sideband power ratio is
sufficiently high so that the contribution of the sideband shot
noise to the overall shot noise is small. Once the Michelson
interferometer and signal-recycling cavity are locked, the Michelson
offset feedback control system is turned on and the offset is
optimized such that the interferometer output power is 100 $\mu$W.
It is also important to optimize the control gain to stabilize the
power so that drift of the shot noise level is sufficiently lower
than the effective squeezing level. The electronic noise of the
readout is 6 dB below the shot noise. The combination of the
detection efficiency (93\%), the squeezing injection efficiency
associated with the injection and round-trip loss (77.5\%), the
mode-overlap efficiency between the squeezed vacuum and
interferometer field (96.0\%), and the interferometer response
determined by the reflectivity of the end test masses (99.5\%), the
reflectivity of the signal-recycling mirror (92.5\%), and the
Michelson offset ($\pi/238$) attenuates the injected squeezing level
of 9.3$ \pm $0.1 dB down to about 3 dB \cite{goda2007mit}. 

The squeezing phase is locked to the amplitude quadrature of the
interferometer field by the noise-locking technique
\cite{mckenzie2005job} using a PZT-actuated mirror. It is critical
to stably control the squeezing phase relative to the interferometer
field since the anti-squeezing would, otherwise, degrade the shot
noise level.



%
%

\end{document}